\documentclass[%
reprint,
superscriptaddress,
 prl,
 amsmath,
 amssymb,
 aps,
 longbibliography
]{revtex4-2}

\usepackage{cancel}
\usepackage{graphicx}
\usepackage{dcolumn}
\usepackage{bm}
\usepackage[usenames,dvipsnames]{color}
\usepackage{xcolor}
\usepackage{soul}
\usepackage{enumitem}
\usepackage{xspace}
\usepackage{booktabs}
\usepackage{lipsum}
\usepackage{siunitx} 
\usepackage[ISO]{diffcoeff}
\usepackage{subcaption}
\usepackage{ragged2e}
\usepackage{mathrsfs}
\usepackage{hyperref}
\usepackage[capitalise]{cleveref}
\usepackage[normalem]{ulem}
\usepackage{amsmath}
\usepackage{esvect} 
\usepackage{titlesec}
\usepackage{titletoc}
\usepackage{chngcntr}
\usepackage{fontawesome5}
\usepackage{mleftright}
\usepackage[nodayofweek]{datetime}
\usepackage{comment}

\newcommand{\Cv}{C_{\!v}}
\newcommand{\FS}{F_S}
\newcommand{\FT}{F_T}
\newcommand{\Fb}{F_\beta}
\newcommand{\kB}{k_{\mathrm{B}}}
\newcommand{\Tr}{\mathrm{Tr}}
\newcommand{\Var}{\mathrm{Var}}
\newcommand{\Cov}{\mathrm{Cov}}

\definecolor{orcid-green}{RGB} {166, 206, 57}
\newcommand{\MYhref}[3][blue]{\href{#2}{\color{#1}{#3}}}%

\begin{document}

\title{Quantum Fisher Information for Entropy of Gibbs States}

\author{Francis J. Headley\,\MYhref[orcid-green]{https://orcid.org/0009-0000-7585-4957}{\faOrcid}} 
\email{francis.headley@uni-tuebingen.de}
\affiliation{Institut f\"{u}r Theoretische Physik,
  Eberhard-Karls-Universit\"{a}t T\"{u}bingen,
  72076 T\"{u}bingen, Germany}
\date{\today}

\begin{abstract}
We derive the quantum Fisher information for entropy estimation in a Gibbs state and show that it equals the inverse of the heat capacity, which is dual to the temperature Fisher information given by the heat capacity divided by the square of the temperature. Their product is independent of the Hamiltonian and depends only on the temperature, leading to a metrological uncertainty relation between the variances of entropy and temperature estimators in which all system-specific quantities cancel. This relation arises from the dually‑flat structure of the Gibbs exponential family expressed in thermodynamic coordinates, and holds for all standard thermodynamically conjugate pairs. We identify energy measurement as the optimal protocol for entropy estimation, analyse critical‑point scaling where the entropy Fisher information vanishes, and connect it to the Ruppeiner metric in entropy coordinates. We lastly examine the distinguished role of the von Neumann entropy within the Rényi family. Generalisations to the grand canonical and generalised Gibbs ensembles are given.
\end{abstract}

\maketitle

\section{Introduction}
The Legendre structure of thermodynamics pairs each intensive variable with a conjugate extensive one: temperature $T$ with entropy $S$, pressure $P$ with volume $V$, chemical potential $\mu$ with particle number $N$. Among these, the pair $(T,S)$ occupies a distinguished role. Temperature governs the statistical weight of microstates, $p_k \propto e^{-E_k/T}$, and temperature estimation from quantum measurements of thermal states has a well-developed metrological theory~\cite{braunstein1994, paris2009, correa2015, mehboudi2019, depasquale2016}. The quantum Fisher information (QFI) for temperature is $\FT = \Cv/T^2$ (we set $\kB = 1$ throughout), yielding the Cram\'{e}r--Rao bound $\Var(\hat{T}) \geq T^2/(n\Cv)$. Yet the question, with what precision can entropy \emph{itself} be inferred from quantum measurements, has not been explicitly presented.
This gap is notable, as entropy is a central quantity not only in theory but also in experiment, where its determination via thermodynamic integration~\cite{nascimbene2010} is a key tool in systems such as ultracold quantum gases and quantum dots~\cite{child2022, hartman2018}. Maxwell relations also provide a complementary route to thermometry: high precision in temperature estimation occurs when the entropy changes rapidly with respect to a control parameter~\cite{mihailescu2023}, and the interplay between thermometric and entropic sensitivity in nano-electronic devices has been explored through the Fisher information~\cite{mihailescu2024}.
In this work we derive the QFI for entropy estimation, which is dual to the temperature QFI, and show that their product yields a Cram\'{e}r--Rao relation whose form reflects the dually-flat geometry of the Gibbs exponential family when expressed in thermodynamic coordinates.

\section{Entropy Fisher information}
Consider $n$ independent copies of a Gibbs state $\rho(\beta) =
e^{-\beta H}/Z$ at inverse temperature $\beta = 1/T$.  Because
$\rho(\beta)$ belongs to an exponential family, its eigenvalues $p_k =
e^{-\beta E_k}/Z$ are determined by $\beta$ alone and the energy
eigenbasis $\{|k\rangle\}$ is $\beta$-independent.  The QFI with
respect to $\beta$ is therefore purely classical~\cite{braunstein1994}:
\begin{equation}\label{eq:Fbeta}
  \Fb = \sum_k \frac{(\partial_\beta p_k)^2}{p_k} = \Var(H)
  = \frac{\Cv}{\beta^2}\,,
\end{equation}
where $\Cv \equiv \beta^2\Var(H)$ is the heat capacity. The von~Neumann entropy $S = -\Tr(\rho\ln\rho) = \beta\langle
H\rangle + \ln Z$ is a smooth, strictly monotone function of $\beta$
($dS/d\beta = -\Cv/\beta < 0$ for any thermodynamically stable
system with $\Cv > 0$).  It therefore provides a valid
reparametrisation of the one-dimensional Gibbs manifold, and the QFI transforms covariantly as
\begin{equation}\label{eq:main}
  \FS = \left(\frac{d\beta}{dS}\right)^{\!2}\Fb = \frac{1}{\Cv}\,.
\end{equation}
The Cram\'{e}r--Rao bound gives
\begin{equation}
    \Var(\hat{S}) \geq \dfrac{1}{n\FS} = \dfrac{\Cv}{n}.
\end{equation}
This bound is saturated asymptotically by a projective energy measurement (see Appendix). Equation~\eqref{eq:main} indicates a duality: the heat capacity governs the precision of $S$- and $T$-estimation inversely. Having a large $\Cv$ allows for precise temperature estimation ($\FT \propto \Cv$) as small temperature differences produce large, distinguishable shifts in the energy distribution.  Conversely, large $\Cv$ makes entropy hard to estimate ($\FS \propto 1/\Cv$) because many distinct entropy values map to nearly identical energy statistics.

\begin{figure*}[t!]
\centering
\includegraphics[width=1.8\columnwidth]{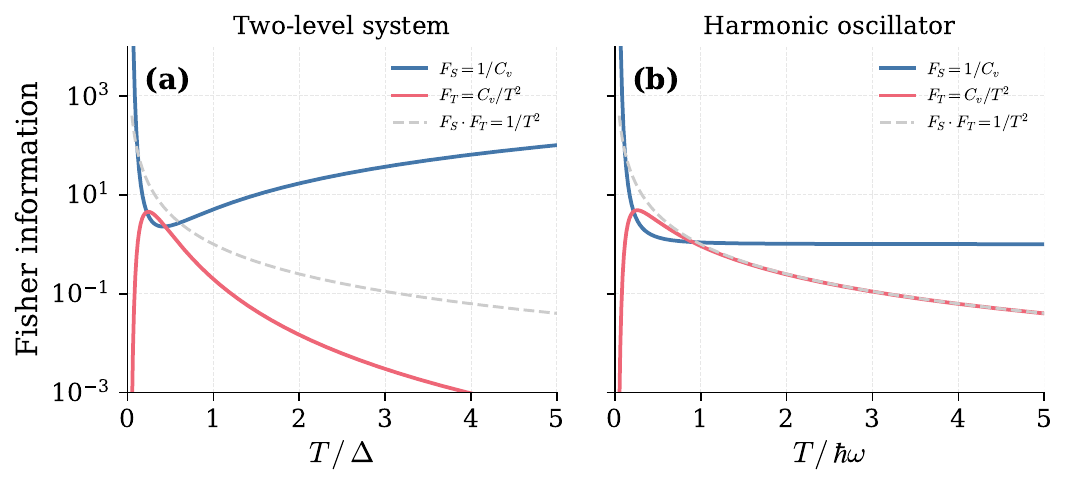}
\caption{Fisher information for entropy and temperature estimation in (a) a two-level system with energy gap $\Delta$ and (b) a quantum harmonic oscillator with frequency $\omega$.  Solid curves show $\FS = 1/\Cv$ (blue) and $\FT = \Cv/T^2$ (red); the dashed curve is their product $\FS \cdot \FT = 1/T^2$.  In panel (b), green dots mark the two crossovers at $\Cv = T$ ($T \approx 0.22\,\hbar\omega$ and $T \approx 0.90\,\hbar\omega$); the shaded region indicates the intermediate ``thawing'' regime where $\FT > \FS$.  The dotted line marks the classical equipartition floor $\FS \to 1$.}
\label{fig:fisher}
\end{figure*}

\section{Entropy--temperature metrological uncertainty relation}

The product of the two Fisher informations is
\begin{equation}\label{eq:product}
  \FS\cdot\FT = \frac{1}{\Cv}\cdot\frac{\Cv}{T^2} = \frac{1}{T^2}\,,
\end{equation}
in which $\Cv$ cancels identically.  Applying the Cram\'{e}r--Rao bounds to both variables for $n$ independent copies gives
\begin{equation}\label{eq:uncertainty}
  \Delta^2 S\;\Delta^2 T \;\geq\; \frac{T^2}{n^2}\,,
\end{equation}
where $\Delta^2 X \equiv \Var(\hat{X})$ denotes the variance of the estimator.  The bound depends only on the temperature of the state; no system-specific quantity appears.  It is saturated when both estimators are constructed from energy measurements on $n$ copies.
Crucially, this bound is saturatable: because projective energy measurement simultaneously saturates both individual Cram\'{e}r--Rao bounds, the product $\Delta^2 S\,\Delta^2 T = T^2/n^2$ is achievable, distinguishing Eq.~\eqref{eq:uncertainty} from a generic (and potentially vacuous) product of independent lower bounds. An analogous relation holds for every Legendre-conjugate pair, reflecting a common origin: for any conjugate pair $(\lambda, A)$ entering the Gibbs weight through a natural parameter $\gamma = \beta\lambda$, the product $F_\lambda \cdot F_A = \beta^2 = 1/T^2$ follows from the duality of natural and expectation parameters in exponential families, with the factor $\beta^2$ arising from the coordinate change $\gamma \mapsto \lambda$.  

\section{Two-level system}

For a two-level system with energy gap $\Delta$, the heat capacity is $\Cv = (\beta\Delta)^2 e^{\beta\Delta}/(1+e^{\beta\Delta})^2$, giving
\begin{equation}
  \FS = \frac{(1+e^{\beta\Delta})^2}{(\beta\Delta)^2\,e^{\beta\Delta}}\,.
\end{equation}
At high temperature ($\beta\Delta \to 0$), $\Cv \to 0$ and $\FS \to \infty$: the state approaches the maximally mixed state ($S = \ln 2$), which is sharply localised in entropy space.  At low temperature ($\beta\Delta \to \infty$), $\Cv \to 0$ exponentially and $\FS \to \infty$ again: the nearly pure ground state has entropy close to zero and is equally well localised.  The minimum of $\FS$---the point of greatest difficulty for entropy estimation---occurs at intermediate temperature where $\Cv$ is largest.  At every temperature, $\FS\cdot\FT = 1/T^2$ holds exactly.

\section{Quantum harmonic oscillator}

For a quantum harmonic oscillator at frequency $\omega$, the heat capacity is
\begin{equation}\label{eq:qho}
  \Cv = \frac{(\beta\hbar\omega)^2\,e^{\beta\hbar\omega}}{(e^{\beta\hbar\omega}-1)^2}\,,
\end{equation}
so $\FS = (e^{\beta\hbar\omega}-1)^2/[(\beta\hbar\omega)^2\,e^{\beta\hbar\omega}]$.  Unlike the two-level system, $\Cv$ is monotonically increasing: the spectrum is unbounded and there is no Schottky anomaly, so the metrological structure is qualitatively different. As $T$ increases, entropy estimation steadily approaches a constant while temperature estimation steadily worsens.

The condition $\FS = \FT$ reduces to $\Cv = T$ , which has two solutions: $T_1 \approx 0.22\,\hbar\omega$ and $T_2 \approx 0.90\,\hbar\omega$. Below ($T \ll T_1$), the nearly pure ground state is sharply localised in entropy space and $\FS$ dominates;  The regime ($T_1 < T < T_2$), $\FT > \FS$, favours thermometry over entropy estimation. In the classical regime ($T \gg T_2$), equipartition sets in and the two informations assume their classical asymptotic values.  
In the classical limit, a system of $f$ quadratic degrees of freedom has $\Cv \to f/2$, yielding:
\begin{equation}
  \FS \;\to\; \frac{2}{f}\,, \qquad \FT \;\to\; \frac{f}{2T^2}\,.
\end{equation}
Each additional degree of freedom dilutes the per-copy information about entropy, making $S$ harder to estimate: the energy distribution broadens as $\Var(H) = fT^2/2$, so a wider range of entropy values produces statistically similar measurements.  The approach to this classical floor is not smooth but reflects the discrete activation of internal modes.  As each degree of freedom unfreezes with increasing temperature, $\Cv$ rises in a characteristic staircase; at the same time, $\FS$ descends in discrete steps, providing a metrological signature of degree-of-freedom activation complementary to the familiar staircase in $\Cv$ (see Appendix).

As for all Gibbs states, the product $\FS \cdot \FT = 1/T^2$ holds at every temperature, independent of $f$.  The number of active degrees of freedom governs how this fixed metrological budget is partitioned between entropy and temperature precision, but cannot alter the budget itself.

\section{Critical-point scaling}

At a second-order phase transition with specific-heat exponent $\alpha$, $\Cv \sim |t|^{-\alpha}$ where $t = (T-T_c)/T_c$.  From Eq.~\eqref{eq:main},
\begin{equation}\label{eq:critical}
  \FS \sim |t|^\alpha \;\to\; 0
  \quad\text{as}\quad t\to 0\,.
\end{equation}
The Cram\'{e}r--Rao bound $\Var(\hat{S}) \geq \Cv/n$ diverges: near criticality, the proliferation of low-energy excitations renders the energy distributions at neighbouring entropy values statistically indistinguishable.  For $\alpha = 0$ (e.g.\ the 2D Ising model~\cite{onsager1944}), the vanishing is logarithmic: $\FS \sim 1/\ln|t|^{-1}$. For a finite system of linear extent $L$, standard scaling gives $\Cv(t,L) \sim L^{\alpha/\nu}\,\tilde{c}(tL^{1/\nu})$, so
\begin{equation}\label{eq:fss}
  \FS(t,L) \sim L^{-\alpha/\nu}\,\tilde{\FS}(tL^{1/\nu})\,.
\end{equation}
At $t=0$ the entropy Fisher information thus decays as a power of system size, providing a direct metrological route to the ratio $\alpha/\nu$: the slope of $\ln\FS$ versus $\ln L$ at criticality equals $-\alpha/\nu$. This behaviour has a dual interpretation.  The thermodynamic length in entropy coordinates~\cite{sivak2012, crooks2007},
\begin{equation}
  \mathcal{L} = \int \frac{dS}{\sqrt{\Cv}}\,,
\end{equation}
also vanishes as $\Cv \to \infty$: entropy-changing protocols become quasi-statically cheap near the critical point.  The divergence of $\Cv$ therefore simultaneously makes entropy hardest to estimate ($\FS \to 0$) and cheapest to change ($\mathcal{L} \to 0$), a complementarity between metrological distinguishability and thermodynamic cost at criticality.

At first-order transitions, the entropy is discontinuous and the Gibbs family breaks down; the QFI framework does not apply across the coexistence region.

\section{Relation to thermodynamic geometry}

Ruppeiner~\cite{ruppeiner1979, ruppeiner1995} defined a Riemannian metric on equilibrium states via $g^R_{UU} = -\partial^2 S/\partial U^2 = 1/(T^2\Cv)$.  Janyszek and Mruga{\l}a~\cite{janyszek1989} showed that this coincides with the Fisher information metric for Gibbs distributions.  Under the coordinate change $dU = T\,dS$, the metric transforms to $g^R_{SS} = 1/\Cv = \FS$.  Thus $\FS$ is precisely the Ruppeiner metric component in entropy coordinates---an identification that is implicit in the information-geometry literature but whose estimation-theoretic content has not previously been exploited.

The QFI framing contributes three elements beyond the geometric picture: (i)~the bound $\Var(\hat{S}) \geq \Cv/n$ is operationally achievable, not merely a measure of statistical distance; (ii)~the optimal measurement (projective energy measurement) is identified explicitly; and (iii)~the metrological uncertainty relation~\eqref{eq:uncertainty} emerges from combining $\FS$ with $\FT$, the metrological implications of this geometric duality have not been previously explored.

\section{Other conjugate pairs} \label{sec:pairs}

The argument of Sec.~II extends to every Legendre-conjugate pair.  In each case the relevant Gibbs state is an exponential family in the intensive parameter, the QFI for that parameter equals $\beta^2$ times the variance of its conjugate operator, and the Fisher information for the extensive variable follows by reparametrisation.  The conjugate susceptibility cancels in the product.

For the $(-P,V)$ pair in the isothermal--isobaric ensemble $\rho = e^{-\beta H - \gamma\hat{V}}/Z$ with $\gamma \equiv \beta P$, one finds $F_P = \beta^2\Var(\hat{V})$ and $F_V = 1/\Var(\hat{V})$, so that
\begin{equation}\label{eq:FV_FP}
  F_V\cdot F_P = \frac{1}{T^2}\,,
\end{equation}
with the isothermal compressibility $\kappa_T$ cancelling identically from the product (see Appendix~\ref{app:pairs} for the full derivation).  The Cram\'{e}r--Rao bound gives $\Delta^2 V\;\Delta^2 P \geq T^2/n^2$.

Similarly, for the $(\mu,N)$ pair in the grand canonical ensemble, $F_\mu = \beta^2\Var(\hat{N})$ and $F_N = 1/\Var(\hat{N})$, yielding
\begin{equation}\label{eq:FN_Fmu}
  F_N\cdot F_\mu = \frac{1}{T^2}\,,
  \qquad
  \Delta^2 N\;\Delta^2\mu \;\geq\; \frac{T^2}{n^2}\,,
\end{equation}
with the particle-number susceptibility $(\partial N/\partial\mu)_T$ cancelling identically.

\begin{table}[h]
\centering
\renewcommand{\arraystretch}{1.3}
\begin{tabular}{lcccc}
\hline\hline
Pair $(\lambda, A)$ & $F_\lambda$ & $F_A$
  & $F_\lambda F_A$ & Bound on $\Delta^2\lambda\,\Delta^2 A$ \\
\hline
$(T,\; S)$
  & $\Cv/T^2$ & $1/\Cv$
  & $1/T^2$ & $T^2/n^2$ \\[2pt]
$(-P,\; V)$
  & $\beta^2\Var(\hat{V})$ & $1/\Var(\hat{V})$
  & $1/T^2$ & $T^2/n^2$ \\[2pt]
$(\mu,\; N)$
  & $\beta^2\Var(\hat{N})$ & $1/\Var(\hat{N})$
  & $1/T^2$ & $T^2/n^2$ \\
\hline\hline
\end{tabular}
\caption{Fisher information relations and metrological uncertainty bounds for the
three standard thermodynamic conjugate pairs.  In every case the
conjugate susceptibility cancels identically from the product and the
bound $T^2/n^2$ holds.}
\label{tab:pairs}
\end{table}

The universality has a common origin: for any conjugate pair $(\lambda, A)$ entering the Gibbs weight through a natural parameter $\gamma = \beta\lambda$, one has $F_\gamma = \Var(\hat{A})$ and $F_A = 1/\Var(\hat{A})$, while the map $\gamma \mapsto \lambda$ introduces a factor $\beta^2$ in $F_\lambda$, yielding $F_\lambda\cdot F_A = \beta^2 = 1/T^2$ in every case.  After restoring $\kB$, the bound becomes $\kB^2 T^2/n^2$.  Table~\ref{tab:pairs} summarises these results.

\section{R\'{e}nyi entropy Fisher information}
\label{sec:renyi}

The von~Neumann entropy $S = S_1$ is distinguished among the R\'{e}nyi family $S_\alpha = (1{-}\alpha)^{-1}\ln\Tr(\rho^\alpha)$ by its direct connection to thermodynamic conjugacy.  For a Gibbs state, one may show (Appendix) that $S_\alpha(\beta)$ is strictly monotone for every $\alpha > 0$, so the reparametrisation argument applies, yielding the R\'{e}nyi entropy Fisher information
\begin{equation}\label{eq:renyi_FS}
  \FS^\alpha
  = \frac{(\alpha{-}1)^2\,\Var_\beta(H)}
         {\alpha^2\bigl[U(\alpha\beta) - U(\beta)\bigr]^2}
  \equiv \frac{1}{\Cv^\alpha}\,,
\end{equation}
where $U(x) \equiv \langle H\rangle_x$ is the mean energy at inverse temperature $x$, and $\Cv^\alpha$ defines a \emph{R\'{e}nyi heat capacity} that reduces to $\Cv$ as $\alpha \to 1$. The product with the thermometric Fisher information is
\begin{equation}\label{eq:renyi_product}
  \FS^\alpha \cdot \FT
  = \frac{(\alpha{-}1)^2\,\Cv^2}
         {\alpha^2\bigl[U(\alpha\beta) - U(\beta)\bigr]^2}\,.
\end{equation}
For $\alpha \neq 1$, this depends on the Hamiltonian through $U(\alpha\beta)$ and does \emph{not} reduce to $1/T^2$.  The universal cancellation of $\Cv$ in $\FS \cdot \FT$ is therefore special to the von~Neumann entropy: it is the unique member of the R\'{e}nyi family for which $dS/d\beta \propto \Cv$, the same quantity that governs $\Fb$.  This singles out $S_1$ as the metrologically natural entropy for Gibbs-state estimation.

In the classical limit ($f$ quadratic degrees of freedom), $U(x) = f/(2x)$ and the R\'{e}nyi Fisher information becomes $\FS^\alpha \to 2/f$, independent of $\alpha$.  This reflects the Gaussian nature of the energy distribution in this regime, where all R\'{e}nyi entropies are affine functions of $S_1$.

\section{Generalisations}

\textit{Grand canonical ensemble.}---For $\rho(\beta,\mu) =
e^{-\beta(H-\mu N)}/\mathcal{Z}$, projecting the two-parameter Fisher
matrix onto the entropy direction (see
Appendix~\ref{app:gce}) yields
\begin{equation}\label{eq:gce}
  \FS^{\mathrm{GCE}} = \frac{1}{\beta^2\Var(H-\mu N)}
  = \frac{1}{\Cv^{(\mu)}}\,,
\end{equation}
where $\Cv^{(\mu)}$ is the heat capacity at constant chemical potential.  When $\langle N\rangle$ is additionally held fixed, $\FS|_{\langle N\rangle} = 1/\Cv^{(N)}$ with $\Cv^{(N)} = \beta^2\!\left(\Var(H) - \Cov(H,N)^2/\Var(N)\right)$.

\textit{Generalised Gibbs ensemble.}---For
$\rho = e^{-\sum_k\lambda_k I_k}/Z$ with mutually commuting conserved charges $\{I_k\}$ and conjugate Lagrange multipliers $\{\lambda_k\}$, the entropy $S = \sum_k\lambda_k\langle I_k\rangle + \ln Z$ can be estimated by projecting the multiparameter Fisher matrix $[F]_{kl} = \Cov(I_k, I_l)$ onto the entropy gradient (see Appendix~\ref{app:gge}), giving
\begin{equation}\label{eq:gge}
  \FS^{(\mathrm{GGE})} = \frac{1}{\bm\lambda^T\!F\,\bm\lambda}
                        = \frac{1}{\Cv^{\mathrm{eff}}}\,.
\end{equation}
In all cases the universal form $\FS = 1/\Cv^{\mathrm{eff}}$ is preserved and the metrological uncertainty relation~\eqref{eq:uncertainty} holds with $\Cv \to \Cv^{\mathrm{eff}}$. When the conserved charges $\{I_k\}$ mutually commute, the SLD operators $L_{\lambda_k} \propto (I_k - \langle I_k\rangle)$ also commute, so the multiparameter SLD bound is simultaneously saturable.  In particular, the optimal measurements for all extensive variables---entropy, volume, particle number---are mutually compatible: a single projective measurement in the joint eigenbasis of the charges saturates every scalar Cram\'{e}r--Rao bound at once.  Incompatibility costs arise only for non-Abelian generalised Gibbs ensembles, where the eigenbasis rotates with the parameters.

\section{Discussion}

The entropy Fisher information $\FS = 1/\Cv$ is dual to the thermometric Fisher information $\FT = \Cv/T^2$, and their product yields the Cram\'{e}r--Rao relation $\Delta^2 S\,\Delta^2 T \geq T^2/n^2$.  The cancellation of system-specific quantities follows from the exponential family structure of Gibbs states; the physical content lies in the duality of estimation directions, the critical-point scaling, and the identification of the optimal measurement.

Meng and Shi~\cite{meng2025} obtained thermodynamic uncertainty relations for parameters encoded in the Hamiltonian, $\hat{H}(\theta)$, with conjugate operator $\hat{O} = \partial_\theta\hat{H}$. The $(T,S)$ pair lies outside that framework: $\beta$ parametrises the Gibbs weight rather than the Hamiltonian, and $S$ is a state functional rather than an observable. The universal product $\FS\cdot\FT = 1/T^2$ and the resulting cancellation of all system-specific quantities follow instead from the reparametrisation properties of the exponential family.

Experimentally, the bound $\Var(\hat{S}) \geq \Cv/n$ is directly relevant to ultracold-atom platforms that determine entropy via thermodynamic integration~\cite{nascimbene2010, trotzky2012}: it sets the irreducible statistical floor on such measurements given $n$ experimental repetitions.  In mesoscopic systems where entropy is accessed via Maxwell relations~\cite{child2022, hartman2018}, the bound constrains the precision achievable from charge-sensing measurements.  The metrological uncertainty relation~\eqref{eq:uncertainty} further implies that simultaneous high-precision determination of both $S$ and $T$ faces a fundamental trade-off governed only by $T$ and $n$.  The connection to fundamental limits of equilibrium metrology~\cite{abiuso2025} suggests broader implications for the design of thermal probes at criticality.

\begin{acknowledgments}
\emph{Acknowledgments} We would like to thank George Mihailescu and Paulo Abiuso for their comments and feedback on the draft. We would like to thank Shubhang Dadhich, Kjartan van Driel, Ned O'Reilly, Marc Nairn, Mathias Bo Svendsen, Parvinder Solanki and particularly Enrique Verduras for fruitful discussions. 
\end{acknowledgments}

\bibliography{refs.bib}

\appendix

\section{Reparametrisation validity}
\label{app:reparam}

The entropy $S(\beta) = \beta\langle H\rangle + \ln Z$ satisfies $dS/d\beta = -\Cv/\beta < 0$ for any stable system ($\Cv > 0$). It is therefore a smooth, strictly decreasing bijection from $(0,\infty)$ to $(S_{\min}, S_{\max})$, establishing a global diffeomorphism between the $\beta$- and $S$-parametrisations of the one-dimensional Gibbs family.  Estimating $S$ is operationally equivalent to estimating $\beta$ and applying the deterministic map $\beta(S)$.

The QFI transforms covariantly under any smooth injective reparametrisation~\cite{paris2009}: for $\beta \mapsto S(\beta)$,
\begin{equation}
  \FS = \left(\frac{d\beta}{dS}\right)^{\!2}\Fb
      = \frac{\beta^2}{\Cv^2}\cdot\frac{\Cv}{\beta^2} = \frac{1}{\Cv}\,.
\end{equation}
This requires: (i)~$\Cv > 0$ (thermodynamic stability), ensuring injectivity of $S(\beta)$; and (ii)~a $\beta$-independent eigenbasis, ensuring $\Fb = \Var(H) = \Cv/\beta^2$ is purely classical.  Both conditions hold for the canonical Gibbs state.

\section{Saturation by energy measurement}
\label{app:saturation}

The symmetric logarithmic derivative (SLD) of $\rho(\beta)$ with
respect to $\beta$ is
\begin{equation}
  L_\beta = \sum_k \frac{\partial_\beta p_k}{p_k}|k\rangle\langle k|
\end{equation}
which is diagonal in the energy eigenbasis.  Under the reparametrisation
$\beta \mapsto S$, the SLD becomes
\begin{equation}
  L_S = \frac{d\beta}{dS}\,L_\beta
\end{equation}
also diagonal in $\{|k\rangle\}$.  The optimal POVM is therefore a projective energy measurement for both $\beta$- and $S$-estimation.

For $n$ independent copies, the sample mean $\bar{E} = n^{-1}\sum_i E_i$ is a sufficient statistic for $\beta$.  The maximum-likelihood estimator $\hat{\beta}_{\mathrm{MLE}}(\bar{E})$ is
asymptotically efficient~\cite{paris2009}, thus
\begin{equation}
  \Var(\hat{S}) \;\xrightarrow{n\to\infty}\;
  \left(\frac{dS}{d\beta}\right)^{\!2}\frac{1}{n\Fb}
  = \frac{\Cv^2/\beta^2}{n\,\Cv/\beta^2} = \frac{\Cv}{n}\,,
\end{equation}
saturating the Cram\'{e}r--Rao bound exactly.

\section{Grand canonical ensemble}
\label{app:gce}
For $\rho(\beta,\mu) = e^{-\beta(H-\mu N)}/\mathcal{Z}$, the
Fisher matrix with respect to $(\beta, \mu)$ has entries
\begin{align}
  F_{\beta\beta}
    &= \Var(H{-}\mu N)\,, \nonumber\\
  F_{\mu\mu}
    &= \beta^2\,\Var(N)\,, \label{eq:gce_fisher}\\
  F_{\beta\mu}
    &= -\beta\,\Cov(H{-}\mu N,\, N)\,. \nonumber
\end{align}
The entropy $S = \beta\langle H-\mu N\rangle + \ln\mathcal{Z}$
depends on both $\beta$ and $\nu = \beta\mu$.  Projecting onto the
entropy direction via the chain rule gives
Eq.~\eqref{eq:gce}~\cite{marzolino2013, hovhannisyan2018}.

\section{Generalised Gibbs ensemble}
\label{app:gge}

For $\rho = e^{-\sum_k\lambda_k I_k}/Z$ with mutually commuting conserved charges $\{I_k\}$, the multiparameter Fisher matrix is $[F]_{kl} = \Cov(I_k,I_l)$, and the entropy is $S = \sum_k\lambda_k\langle I_k\rangle + \ln Z$.  Direct differentiation of $\ln Z$ gives $\partial_{\lambda_k}\ln Z = -\langle I_k\rangle$, so $\partial_{\lambda_k} S = \lambda_k[F]_{kk} + \sum_{l\neq k}\lambda_l[F]_{kl}$, i.e.\ $\nabla_{\bm\lambda} S =- F\bm\lambda$.  The scalar Fisher information for $S$ is obtained by the multiparameter chain rule~\cite{paris2009},
\begin{equation}
  \FS^{(\mathrm{GGE})}
  = [(\nabla_{\bm\lambda} S)^T F^{-1} (\nabla_{\bm\lambda} S)]^{-1}
  = (\bm\lambda^T F\,\bm\lambda)^{-1}
  = \frac{1}{\Cv^{\mathrm{eff}}}\,,
\end{equation}
where $\Cv^{\mathrm{eff}} \equiv (\bm\lambda^TF\bm\lambda)^{-1}$ defines the effective heat capacity of the GGE.

\section{Metrological equipartition in the classical limit}
\label{app:equipartition}

Consider a collection of oscillators with $f$ total degrees of freedom, $H = \sum_{i=1}^{f} \tfrac{1}{2}\kappa_i \xi_i^2$, where the $\xi_i$ are generalised coordinates or momenta with stiffnesses $\kappa_i$.  In the classical limit $T \gg \max_i(\hbar\omega_i)$, each quadratic degree of freedom contributes $\tfrac{1}{2}\kB T$ to the mean energy and $\tfrac{1}{2}$ to the heat capacity, giving
\begin{equation}\label{eq:equip_Cv}
  \Cv \;\to\; \frac{f}{2}\,, \qquad
  \FS \;\to\; \frac{2}{f}\,, \qquad
  \FT \;\to\; \frac{f}{2T^2}\,.
\end{equation}
The entropy Fisher information per degree of freedom is $\FS/f = 2/f^2$, meaning each additional quadratic mode makes entropy harder to estimate by diluting the per-copy information content.  This has a transparent physical origin: with more degrees of freedom the energy distribution broadens ($\Var(H) = f T^2/2$), so a wider range of entropy values produces statistically similar energy measurements.

The approach to this classical floor is governed by the quantum corrections.  For a single oscillator at frequency $\omega$, expanding $\Cv$ to next order gives $\Cv = 1 - (\beta\hbar\omega)^2/12 + \mathcal{O}(\beta^4)$, so
\begin{equation}\label{eq:FS_classical_corr}
  \FS = 1 + \frac{(\beta\hbar\omega)^2}{12} + \mathcal{O}(\beta^4)\,.
\end{equation}
The leading quantum correction \emph{increases} $\FS$: as the temperature drops below the classical regime, the freezing-out of high-energy states begins to constrain the entropy, making it slightly easier to estimate.  For a collection of $f$ oscillators with frequencies $\{\omega_i\}$,
\begin{equation}\label{eq:FS_multi}
  \FS = \frac{2}{f}\left(1 + \frac{\sum_i (\beta\hbar\omega_i)^2}{6f}
        + \mathcal{O}(\beta^4)\right).
\end{equation}
The quantum correction is weighted by the mean-square frequency, $\overline{\omega^2} = f^{-1}\sum_i\omega_i^2$.  Systems with a broad frequency spectrum (e.g.\ a Debye solid) therefore depart from the equipartition floor at higher temperatures than systems with a narrow spectrum, and the departure is anisotropic in frequency space.

For a three-dimensional monatomic ideal gas ($f = 3$ translational modes), $\Cv \to 3/2$ and $\FS \to 2/3$.  A diatomic gas at temperatures above the rotational but below the vibrational threshold has $f = 5$ (three translations, two rotations) giving $\FS \to 2/5$; upon activating the vibrational mode, $f \to 7$ and $\FS \to 2/7$.  Each activation of a new degree of freedom produces a discrete drop in $\FS$, providing a metrological signature of the stepwise unfreezing of internal modes that is complementary to the familiar staircase in $\Cv$.

The product relation is of course maintained throughout: $\FS \cdot \FT = (2/f)(f/2T^2) = 1/T^2$, independent of $f$.  The number of degrees of freedom governs how the fixed metrological budget $1/T^2$ is partitioned between entropy and temperature precision, but cannot alter the budget itself.

\section{Other conjugate pairs: detailed derivations}
\label{app:pairs}
 
\subsection*{The $(-P, V)$ pair}
 
In the isothermal--isobaric ensemble
$\rho = e^{-\beta H - \gamma\hat{V}}/Z$ with $\gamma \equiv \beta P$
and $\hat{V}$ the volume operator.  At fixed $\beta$, this is an
exponential family in $\gamma$ with a $\gamma$-independent eigenbasis,
so
\begin{equation}
  F_\gamma = \Var(\hat{V})\,.
\end{equation}
Since $d\gamma/dP = \beta$,
\begin{equation}\label{eq:FP}
  F_P = \beta^2\Var(\hat{V})\,.
\end{equation}
The fluctuation--response theorem gives
$\Var(\hat{V}) = T\kappa_T V$, where
$\kappa_T = -(1/V)(\partial V/\partial P)_T > 0$ is the isothermal
compressibility.  The equilibrium volume $\langle\hat{V}\rangle(P)$ is
strictly monotone (stability requires $\kappa_T > 0$), so we may
reparametrise $P \mapsto V$:
\begin{equation}\label{eq:FV}
  F_V = \left(\frac{dP}{dV}\right)^{\!2} F_P
      = \frac{\beta^2\Var(\hat{V})}{(\kappa_T V)^2}\,.
\end{equation}
Substituting $\kappa_T V = \beta\Var(\hat{V})$ (from the
fluctuation--response relation $\Var(\hat{V}) = \kappa_T V/\beta$)
into the denominator gives $(\kappa_T V)^2 = \beta^2\Var(\hat{V})^2$,
so
\begin{equation}\label{eq:FV_final}
  F_V = \frac{\beta^2\Var(\hat{V})}{\beta^2\Var(\hat{V})^2}
      = \frac{1}{\Var(\hat{V})}\,.
\end{equation}
The product is
\begin{equation}
  F_V\cdot F_P = \frac{1}{\Var(\hat{V})}\cdot\beta^2\Var(\hat{V})
               = \beta^2 = \frac{1}{T^2}\,,
\end{equation}
and the Cram\'{e}r--Rao bound gives
\begin{equation}\label{eq:VP-bound}
  \Delta^2 V\;\Delta^2 P \;\geq\; \frac{T^2}{n^2}\,.
\end{equation}
The compressibility $\kappa_T$ cancels identically.

\subsection*{The $(\mu, N)$ pair}
 
In the grand canonical ensemble
$\rho = e^{-\beta(H-\mu\hat{N})}/Z$ with $\nu \equiv \beta\mu$, the
same argument at fixed $\beta$ gives
\begin{equation}\label{eq:Fmu}
  F_\mu = \beta^2\Var(\hat{N})\,,
\end{equation}
and, using
$\Var(\hat{N}) = (\partial N/\partial\mu)_T/\beta$ and strict
monotonicity of $\langle\hat{N}\rangle(\mu)$,
\begin{equation}\label{eq:FN}
  F_N = \frac{1}{\Var(\hat{N})}\,.
\end{equation}
The product $F_N\cdot F_\mu = \beta^2 = 1/T^2$, and the bound is
\begin{equation}\label{eq:Nmu-bound}
  \Delta^2 N\;\Delta^2\mu \;\geq\; \frac{T^2}{n^2}\,.
\end{equation}
The particle-number susceptibility $(\partial N/\partial\mu)_T$
cancels identically.

\section{R\'{e}nyi entropy Fisher information: detailed derivation}
\label{app:renyi}
 
For a Gibbs state $\rho(\beta) = e^{-\beta H}/Z(\beta)$, the
$\alpha$-power trace evaluates to
\begin{equation}\label{eq:renyi_trace}
  \Tr(\rho^\alpha) = \frac{Z(\alpha\beta)}{Z(\beta)^\alpha}\,,
\end{equation}
so
\begin{equation}\label{eq:renyi_S}
  S_\alpha(\beta) = \frac{1}{1{-}\alpha}
    \bigl[\ln Z(\alpha\beta) - \alpha\ln Z(\beta)\bigr]\,.
\end{equation}
Differentiating with respect to $\beta$ and using
$d\ln Z(x)/dx = -U(x)$, where $U(x) \equiv \langle H\rangle_x$ is
the mean energy at inverse temperature $x$, yields
\begin{equation}\label{eq:renyi_dSdb}
  \frac{dS_\alpha}{d\beta}
  = \frac{\alpha}{\alpha{-}1}
    \bigl[U(\alpha\beta) - U(\beta)\bigr]\,.
\end{equation}
For any $\alpha > 0$ with $\alpha \neq 1$ and any thermodynamically
stable system, $U(x)$ is strictly decreasing (since
$dU/dx = -\Var_x(H) < 0$), so the factor in brackets is non-zero and
$S_\alpha(\beta)$ is strictly monotone: a valid reparametrisation
of the Gibbs manifold for every $\alpha$.
 
The R\'{e}nyi entropy Fisher information follows by the same
covariant transformation used in Sec.~II, yielding Eq.~\eqref{eq:renyi_FS}.

\end{document}